\documentclass[aps,floatfix,twocolumn]{revtex4}
\usepackage{graphicx}
\usepackage{epstopdf}
\usepackage{dcolumn}
\usepackage{bm}
\usepackage{setspace}
\usepackage{psfrag}

\newcommand{\beq}{\begin{eqnarray}}
\newcommand{\eeq}{\end{eqnarray}}

\begin{document}
\title{Multi-channel Electronic and Vibrational Dynamics in Polyatomic High-order Harmonic Generation}

\author{A. Ferr\'e$^1$}
\author{A. E. Boguslavskiy$^{2,6}$}
\author{M. Dagan$^3$}
\author{V. Blanchet$^1$}
\author{B. Bruner$^3$}
\author{F. Burgy$^1$}
\author{A. Camper$^4$}
\author{D. Descamps$^1$}
\author{B. Fabre$^1$}
\author{N. Fedorov$^1$}
\author{J. Gaudin$^1$}
\author{G. Geoffroy$^1$}
\author{J. Mikosch$^{2,8}$}
\author{S. Patchkovskii$^{2,8}$}
\author{S. Petit$^1$}
\author{T. Ruchon$^4$}
\author{H. Soifer$^3$}
\author{D. Staedter$^5$}
\author{I. Wilkinson$^2$}
\author{A. Stolow$^{2,6,7}$}
\author{N. Dudovich$^3$}
\author{Y. Mairesse$^1$}
\affiliation{$^1$ Universit\'e de Bordeaux - CNRS - CEA, CELIA, UMR5107, F33405 Talence, France}
\affiliation{$^2$ National Research Council Canada, 100 Sussex Dr, Ottawa, ON K1A 0R6, Canada}
\affiliation{$^3$ Department of Physics of Complex Systems, Weizmann Institute of Science, Rehovot 76100, Israel}
\affiliation{$^4$ CEA, IRAMIS, Lasers, Interactions and Dynamics Laboratory - LIDyL, CEA-SACLAY , F-91191 Gif-sur-Yvette, France}
\affiliation{$^5$ Universit\'e de Toulouse -  CNRS, LCAR-IRSAMC, Toulouse, France }
\affiliation{$^6$ Department of Physics, University of Ottawa. 150 Louis Pasteur ON K1N 6N5, Canada}
\affiliation{$^7$ Department of Chemistry, University of Ottawa. 10 Marie Curie ON K1N 6N5, Canada}
\affiliation{$^8$ Max-Born-Institute, Max-Born-Strasse 2A, 12489 Berlin, Germany}

\date{\today}

\begin{abstract}

High-order harmonic generation in polyatomic molecules generally involves multiple channels, each associated with a different orbital. Their unambiguous assignment remains a major challenge for high-harmonic spectroscopy. Here we present a multi-modal approach, using unaligned SF$_6$ molecules as an example, where we combine methods from extreme-ultraviolet spectroscopy, above-threshold ionization and attosecond metrology. Channel-resolved above-threshold ionization measurements reveal that strong-field ionization opens four channels. Two of them, identified by monitoring the ellipticity dependence of the high-harmonic emission, are found to dominate the harmonic signal. A switch from the HOMO-3 to the HOMO-1 channel at 26 eV is confirmed by a phase jump in the harmonic emission, and by the differing dynamical responses to molecular vibrations. This study demonstrates a \textit{modus operandi} for extending high-harmonic spectroscopy to polyatomic molecules, where complex attosecond dynamics are expected.

\end{abstract}
\maketitle

In the simplest view of molecular strong-field ionization, the most probable valence electron to be removed by a strong laser field is the most weakly bound one, namely that from the highest occupied molecular orbital (HOMO). This assumption is based on quasi-static, single active electron pictures and the exponential decay of the tunnel ionization probability through a field-suppressed Coulomb barrier \cite{keldysh65}. However, even in a quasi-static picture, the tunnel ionization probability also depends on the spatial structure of the molecular orbital \cite{muth-bohm00,tong02}. Furthermore, more than one electron could get driven coherently by the laser field. The time scales of electron motions in molecules may not be fast compared to the laser period and the adiabatic approximation inherent to the quasi-static picture may fail: non-adiabatic multi-electron (NME) dynamics can ensue \cite{Lezius01}. Consequently, strong field ionization of molecules is significantly more complex than from atoms, 
with possible contributions from multiple channels, leaving the ion in a 
coherent superposition of electronic states. Such ionic excitation can result in multi-electron processes and ultra-fast charge migrations on an attosecond timescale \cite{smirnova09,haessler10}.

\begin{figure}
\begin{center}
\includegraphics[width=0.5\textwidth]{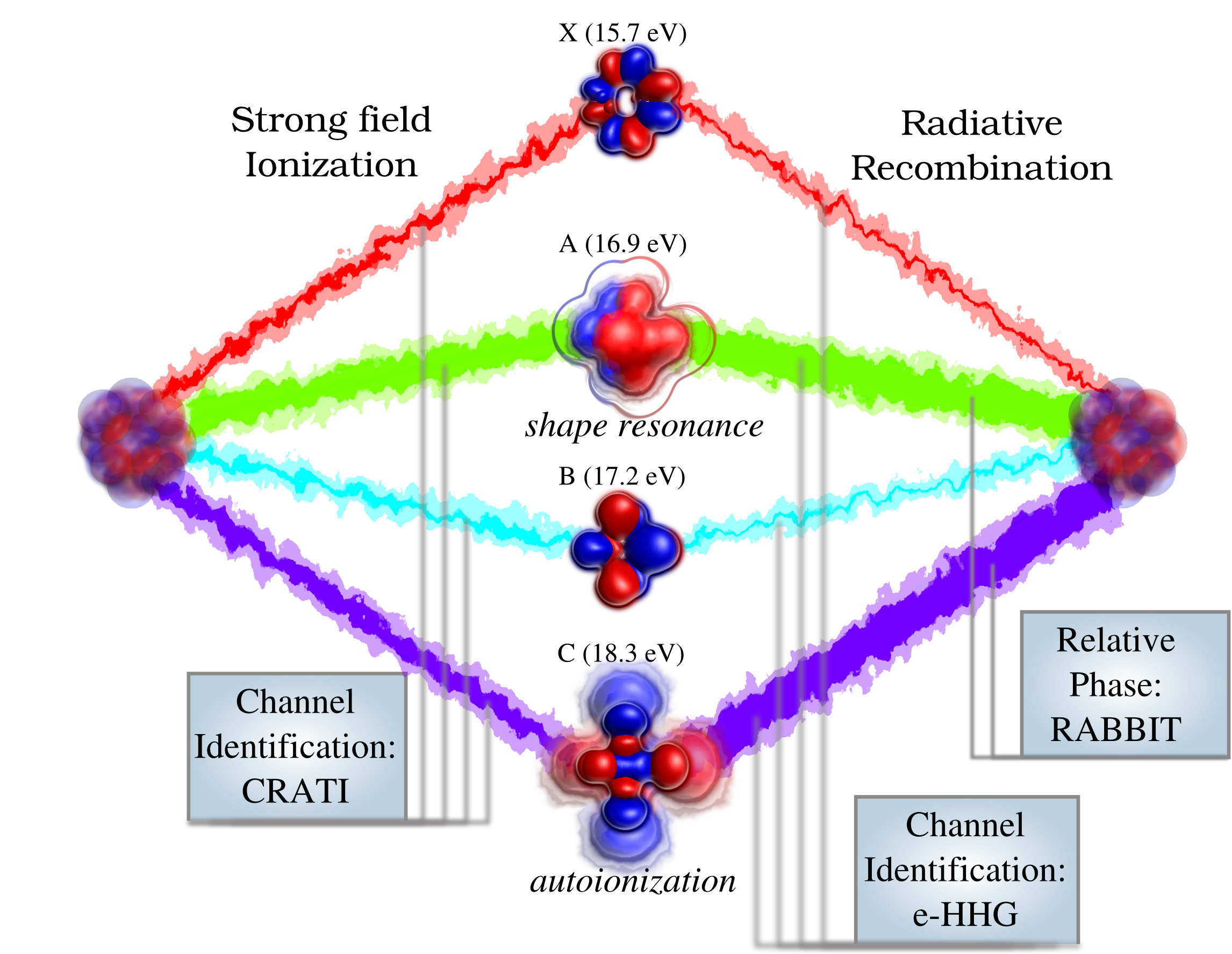}
\caption{Schematic view of multi-channel HHG. Multiple open channels in strong-field ionization are identified with the Channel-Resolved ATI (CRATI) technique. The ionization potential of each channel is indicated next to the schematic view of the associated molecular orbitals. The radiative recombination process is modified by the presence of resonances, increasing some of the channels' contribution to the harmonic emission. These dominant channels are identified by elliptically-driven HHG (e-HHG). The relative phase between the dominant channels is measured by RABBIT. }
\label{Fig0}
\end{center}
\end{figure}

The signatures of multi-channel ionization in above threshold ionization (ATI) \cite{agostini79} and high harmonic generation (HHG) \cite{mcpherson87,corkum93,krause92} are fundamentally different. In ATI, different channels produce different electron energy combs. The relative energy shifts of these combs depend on both the ponderomotive energy, reflecting the laser intensity, and the vertical ionization potential of the channel in question. The contributions of different channels, each leading to an energy-shifted ATI comb, are incoherently summed in the detected photoelectron signal. In contrast, in HHG, the different ionization channels connect the same initial and final states (\textit{i.e.} neutral ground electronic state) and therefore their contributions are coherently summed in the harmonic emission (Fig. \ref{Fig0}). The HHG mechanism can thus be described as a multi-channel interferometer. The weights of the interfering paths are determined by the ionization yields and the recombination dipole 
moments \cite{lewenstein94,le09}. Resonances in the recombination process (e.g. due to the presence of an auto-ionizing state or a shape resonance) strongly increase these dipole moments in specific energy ranges. The relative phases of the interfering paths are complex and have important contributions from all steps of the process. Hence, the HHG mechanism can serve as an extremely sensitive probe which coherently encodes the contribution of the different quantum paths involved in the process -- providing rich spectroscopic signatures of the strong field light matter interaction.

Since the contribution of small amplitude channels is enhanced by the interferometric nature of the process, high-harmonic spectroscopy has the potential to resolve strong field multi-electron phenomena with extremely high sensitivity. However, this task imposes significant challenges. To date, the identification of channels in HHG has been demonstrated in small linear (diatomic or triatomic) molecules via the manipulation of the molecular alignment angle with respect to the generating laser polarization \cite{smirnova09,mcfarland09,haessler10,mairesse10,shafir12}. 
For polyatomic molecules, field-free 3D alignment is achieveable \cite{lee06} but remains challenging in general. Therefore we propose a more direct control over the multi-channel interferometer, illustrated in Fig. \ref{Fig0}. Can we control the relative contributions from each arm? Can we identify the phases associated with the different paths? In the following, we directly address these issues. 

In this letter, we combine complementary strong field measurements to develop a \textit{multi-modal} high-harmonic spectroscopy. To illustrate this, we chose a polyatomic system in which many channels can contribute to the harmonic emission: SF$_6$, whose four highest occupied molecular orbitals (each multiply degenerate) lie within less than 3 eV. Due to its symmetry (O$_h$ point group), this system cannot be aligned in the laboratory frame. The identification of the channels contributing to HHG and the measurement of multi-channel ionization dynamics is thus particularly challenging.

In order to discern the ionization mechanism, we performed channel-resolved ATI (CRATI) measurements on SF$_6$ \cite{boguslavskiy12} for laser wavelengths ranging from 800 to 400 nm, revealing that at least four channels contribute to the strong-field ionization (Fig. \ref{Fig0}). Using high-harmonic spectroscopy, we introduce a new experimental method for identifying  channels based on HHG with elliptically polarized light. The analysis of this reveals a clear transition between two dominant channels around 26 eV. Furthermore, we used RABBIT measurements (Resolution of Attosecond Beating By Interference of Two-photon transitions \cite{paul01}) to reveal the possible influence of resonances in the HHG process \cite{haessler13}. Finally we induced vibrational dynamics in the ground electronic state  of SF$_6$ (via stimulated Raman pumping) in order to explore the influence of molecular geometry on the different channels. The results of these experiments, corroborated by state-of-the-art ab-initio calculations,
 
allowed us to identify the HOMO-1 and HOMO-3 ionization channels as the major contributors to HHG in SF$_6$.

\section{Results}
\subsection{Strong-field ionization channels in SF$_6$}
ATI and HHG are closely related sub-cycle ionization phenomena. In ATI, strong field ionization (SFI) releases electrons into the continuum around every maximum of the laser electric field. This is repeated every optical cycle throughout the pulse. As a result, the photo-electron spectrum shows discrete peaks separated by the driving laser frequency $\omega$. The position of the peaks of this ATI comb is set by both the ionization potential and intensity dependent terms (channel-dependent Stark shifts, ponderomotive shifts U$_p$): KE$_{e^-}^j$=n$\hbar\omega$-I$_p^{j(s)}$-U$_p$, where I$_p^{j(s)}$ is the Stark shifted ($s$) ionization potential of ionization channel $j$, and n is an integer. At constant intensity, the ATI combs of different channels  give a direct measurement of their relative ionization potentials. In polyatomic molecules, where electronically excited cation states tend to undergo field-free fragmentation, different ionization channels can yield different fragments. This means that measuring 
the ATI spectrum associated with a given fragment, using covariance or coincidence detection, provides direct identification of the SFI channel or channels.  

\begin{figure}
\begin{center}
\includegraphics[width=0.5\textwidth]{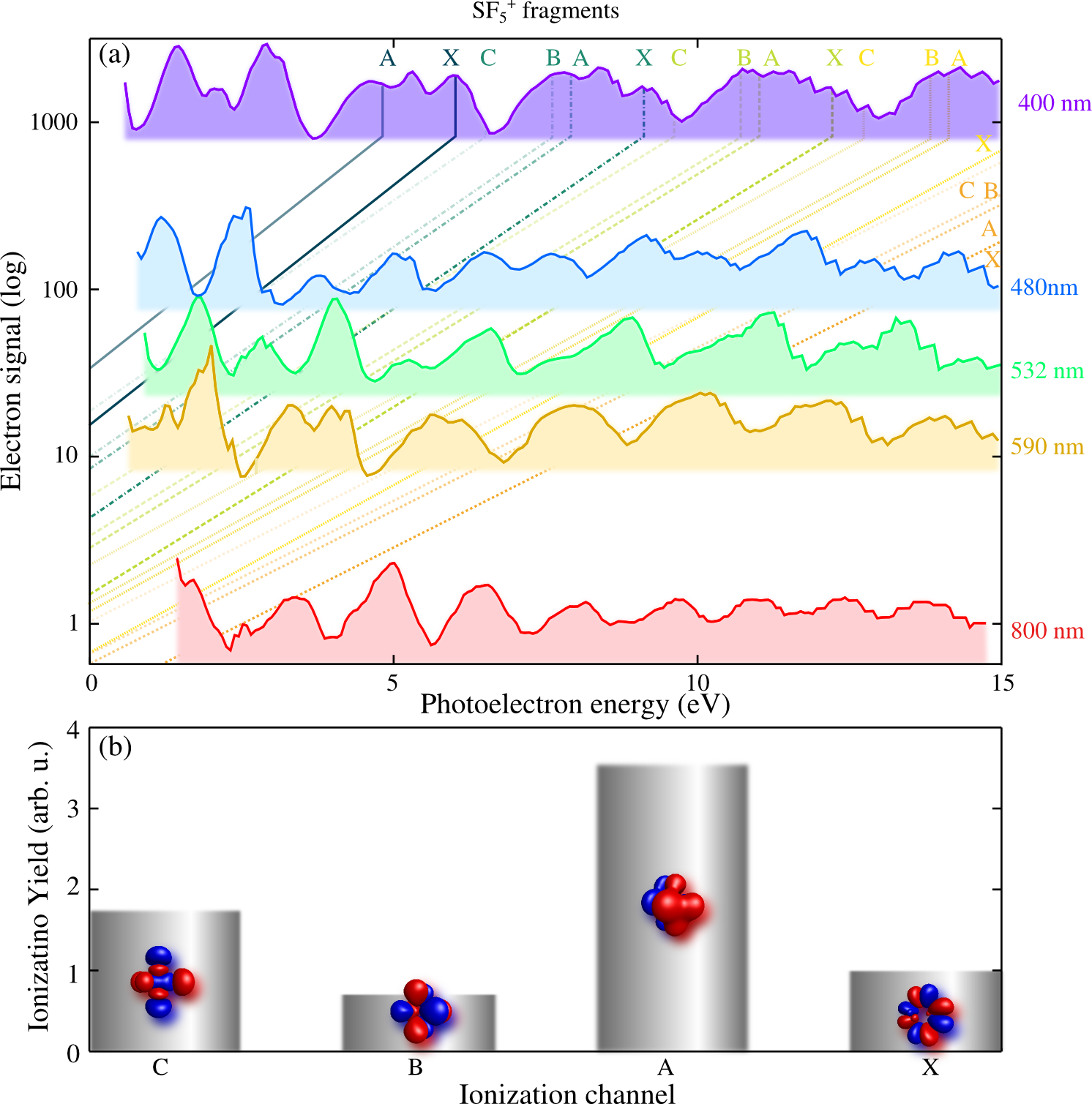}
\caption{(a) Channel-Resolved Above Threshold Ionization (CRATI) spectra of SF$_6$ at different laser wavelengths. Note that the spectra correspond to the SF$_5^+$ fragment; the ground state of SF$_6^+$ is unstable. The spectra are shifted vertically proportional to the laser photon energy. The diagonal lines indicate the expected position of ATI peaks from different ionization channels, for absorption of 7 (dark blue), 8 (blue), 9 (green), 10 (yellow) and 11 (orange) photons. (b) Calculated ionization yield for different channels at 800 nm. The X, A, B and C states are the ground and first three electronically (doublet) excited states of the cation.}
\label{FigCRATI}
\end{center}
\end{figure}

The CRATI technique, however, measures the channel I$_p$ modulo the photon energy. In general, if the energy difference between two ionization channels matches the photon energy of the driving laser, the ATI combs will be overlapping. More specificallyn, if the frequency of the driving laser field is resonant with a single- or multi-photon transition in the neutral or the ion, the ATI combs will also be overlapping. Therefore, in order to resolve the SFI channels, we varied the laser wavelengths between 400 and 800 nm (see Methods). This enabled us to follow the position of the ATI combs as a function of wavelength. The results are shown in Fig. \ref{FigCRATI}, where the ATI spectra associated with the dominant SF$_5^+$ fragment are shown for each laser wavelength. The diagonal lines indicate the expected position of the ATI peaks associated with the four lowest electronic states of the ion, for absorption of 7 to 11 photons. Clearly, multiple channels are present at all wavelengths. However, for the 800 nm 
driving field, the multi-channel nature of the strong-field ionization process is obscured by the cation electronic states having energy level spacings which happen to be multiples of the driving field frequency. Therefore, for the case of SF$_6$, the variation of the driving field frequency was required in order to resolve the multiple SFI channels involved. 

The CRATI measurements show that four main channels are involved in strong field ionization of SF$_6$, at all wavelengths. Quantifying the relative weight of these different channels would require knowledge of the branching ratios to each associated fragment. Unfortunately, these are unknown. In order to theoretically determine these weights, we calculated the tunnel ionization probability $P_{ion}^j$ for each channel $j$ by using a previously described approach \cite{spanner-pra,spanner13} and averaging over all molecular orientations. Ionization calculations used a 1/2-cycle 800-nm sin$^2$ pulse at 10$^{14}$ W/cm$^2$ peak intensity. The resulting weightings are shown in Fig. \ref{FigCRATI} (b). Two main factors determine the channel SFI rates: (i) the ionization potential of the channel -- the ionization rate decreases exponentially with $I_p$; (ii) the topography of the ionizing molecular orbital -- nodes in the plane perpendicular to the laser polarization tend to suppress the ionization probability. The 
first effect was long considered to be dominant: SFI mainly occurs from the highest occupied molecular orbital, even though deeper channels can be observed under specific conditions, such as in aligned molecules. However, in a CRATI study of n-butane \cite{boguslavskiy12}, it was shown that excited state channels dominate over the ground state channel, despite their higher I$_p$s.  The results from Fig. \ref{FigCRATI}(b) are alike due to geometrical effects in SFI of SF$_6$ which overcome the exponential dependence on $I_p$, even for unaligned targets.

\subsection{Calculation of channel-resolved high-harmonic generation}
SFI involving multiple channels generates multiple interfering pathways in high-order harmonic generation. The harmonic signal is determined by the coherent sum over all possible channels (see Fig. 1). In the following, we present a simple model of HHG in which the SFI channels are considered to be uncoupled. Specifically, we assume that there are no sub-cycle non-adiabatic multi-electron transitions (NME) which would coherently couple cation states: an electron leaving a given cation electronic state will recombine to the same state.

\begin{figure}
\begin{center}
\includegraphics[width=0.5\textwidth]{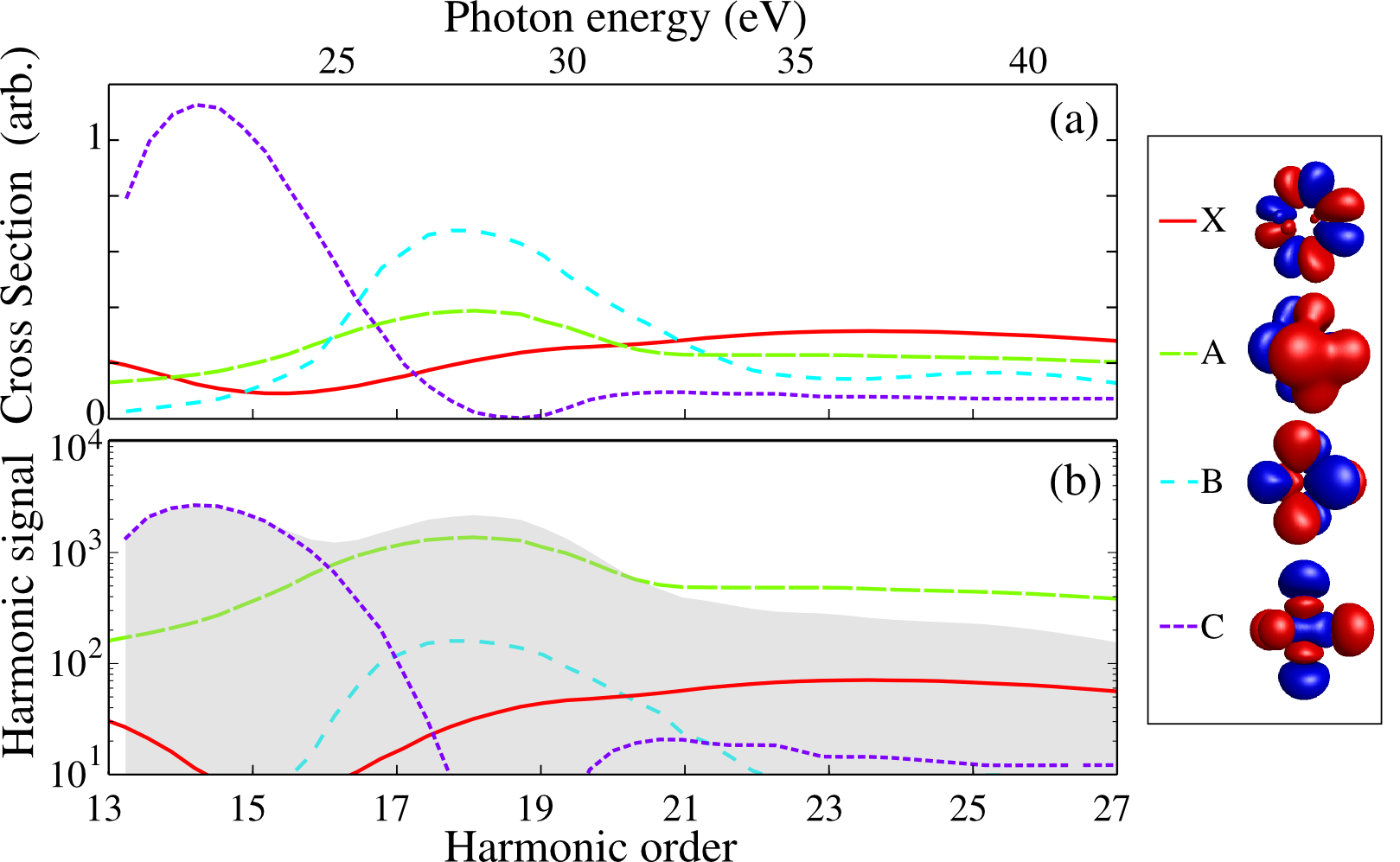}
\caption{(a) XUV photoionization cross sections from \cite{yang98}. (b) Calculated harmonic intensity from different channels, and net response obtained by coherent superposition of all channels (gray area). The inset at right shows the highest occupied molecular orbitals corresponding to the different ionization channels. }
\label{FigTheory}
\end{center}
\end{figure}

The harmonic dipole moment is determined by the SFI probability, the propagation of the electron in the continuum, and the recombination cross section \cite{lewenstein94,le09}. The propagation in the continuum induces a dephasing of the harmonic emission, corresponding to the phase accumulated by the molecule between ionization and recombination. This is given by $\varphi^j=\Delta I_p^j \tau$ \cite{kanai07-2,smirnova09-1}, where $\Delta I_p^j$ is the difference between the ionization potential of the channel under consideration and the ground state ionization channel (X), and $\tau$ is the electron propagation time in the continuum, which increases with harmonic order from 0.6 fs to 1.7 fs for short trajectories at 800 nm. We make the approximation  that the recombination dipole moment $d_{rec}^j$ is equal to the field free XUV photoionization dipole moment. This was studied experimentally using synchrotron radiation and was the subject of accurate theoretical modelling \cite{yang98}. The simulated 
photoionization cross sections from \cite{yang98} are shown in Fig. \ref{FigTheory}(a).

The harmonic intensity  $I^j=\left|P_{ion}^j d_{rec}^j e^{i\varphi^j}\right|^2$, calculated for each channel, is shown in Fig. \ref{FigTheory}(b). The picture, which is very complex in the case of photoionization where many channels have similar cross sections (Fig. \ref{FigTheory}(a)), appears  to be  simplified by SFI. The tunneling probability from the highest occupied molecular orbitals is significantly reduced by the presence of nodes in the orbitals, such that deeper channels are favoured, as illustrated in Fig. \ref{FigCRATI}(b). As a result, the harmonic signal is dominated by two channels, C and A, respectively below and above 25 eV. This property is quite remarkable: due to its sensitivity to tunnel ionization, high harmonic spectroscopy can isolate channels which may be difficult to distinguish in photoionization experiments (for instance, the A and B channels are not distinguished in XUV ionization). The maxima of the dominant channels are related to resonances in the photoionization cross 
section: an autoionization resonance for channel C (5a$_{1g}\rightarrow$6t$_{1u}$) and a shape resonance (1t$_{2u}\rightarrow\epsilon$t$_{2g}$) for channel A \cite{holland92,stener06}.

The total harmonic intensity is calculated by coherently summing the signals from the different channels (Fig. \ref{FigTheory}(b)). The result shows a minimum between H15 and H17 of the 800 nm, around 25 eV. Changing the laser intensity modifies the electron propagation times in the continuum and therefore the relative phase $\varphi^j$ between channels. As a result, the shape and contrast of the minimum are modified, but its position is hardly affected.

\subsection{Experimental determination of SFI  channels}

\begin{figure}
\includegraphics[width=0.4\textwidth]{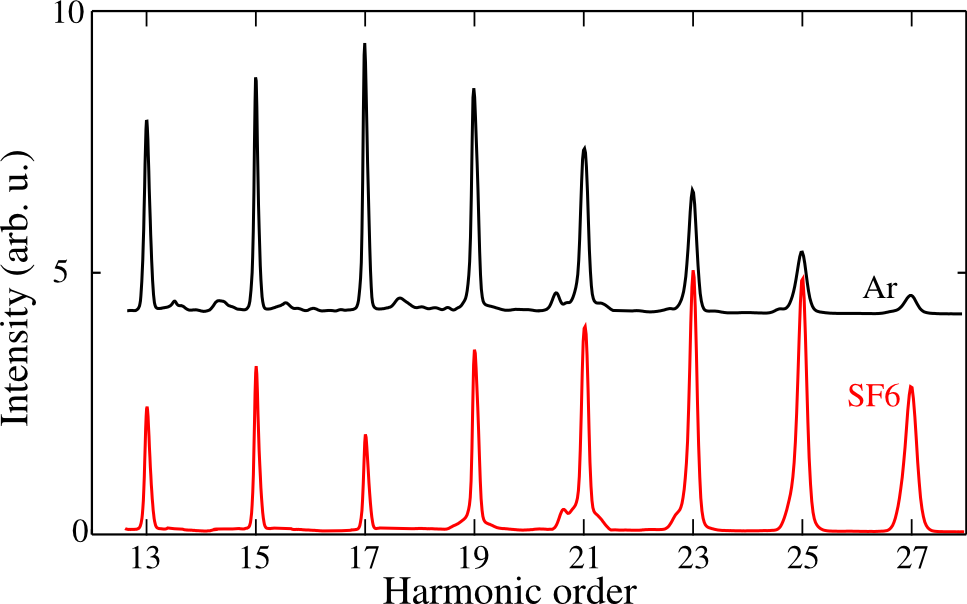}
\caption{Harmonic spectra generated in Ar (black) and SF$_6$ (red) at $I\approx1.3 \times 10^{14}$ W/cm$^2$. The argon spectrum is vertically shifted in order to improve visibility. }
\label{FigSpectrum}
\end{figure}

\begin{figure}
\begin{center}
\includegraphics[width=0.5\textwidth]{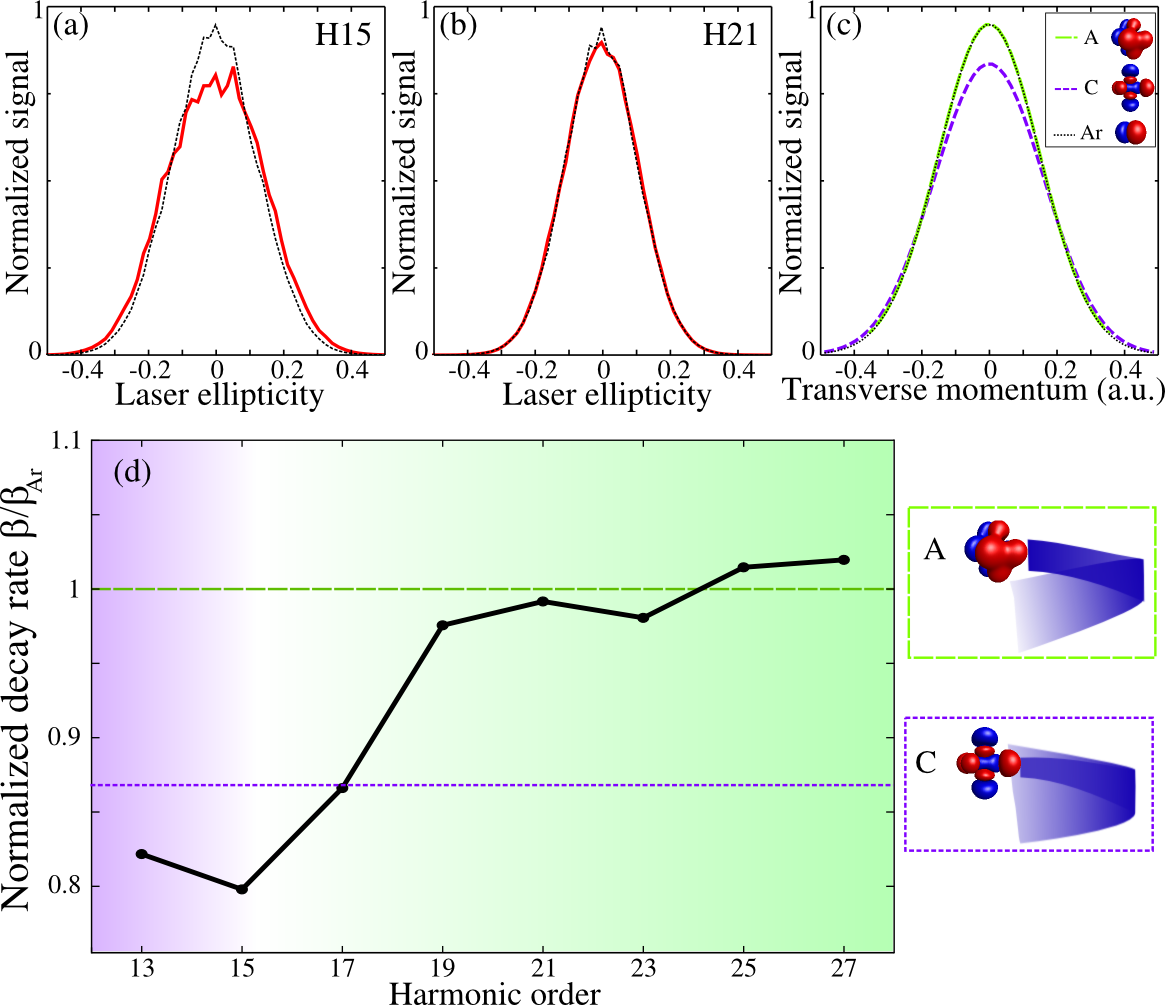}
\caption{The identification of SFI channels using elliptically polarized HHG. Measured harmonic signals as a function of laser ellipticity for H15 (a) and H21 (b) in argon (black dashed) and SF$_6$ (red continuous). The curves are normalized to the same area. (c) Calculated electron transverse momentum distribution in argon and the A and C channel from SF$_6$. (d) Measured response of the harmonic signal as a function of laser ellipticity in SF$_6$, normalized to that in Ar. The horizontal dashed lines are the theoretical normalized ellipticity-dependent responses for the A (green) and C (purple) channels, respectively. }
\label{FigEllipticity}
\end{center}
\end{figure}

Signatures of multi-channel contributions are readily observed in the HHG spectrum. In Figure \ref{FigSpectrum}, we show an experimental harmonic spectrum generated in SF$_6$ using 800 nm pulses at $I\approx1.3 \times 10^{14}$ W/cm$^2$. In order to distinguish the simple, single-orbital picture from the multi-channel case, we compared our measurement with argon, an atomic system with a similar ionization potential to the HOMO of SF$_6$. The SF$_6$ spectrum extends to higher orders and shows a pronounced minimum at harmonic 17, as was previously observed at a lower laser intensity \cite{lynga96}. Varying the laser intensity in our experiment does not modify the position of this minimum, as expected from our simple model. This minimum is a first indication that points towards a transition between channels C and A at around 25 eV (Fig. \ref{FigTheory} (b)). However, a definite identification of such a transition requires new methods in high-harmonic spectroscopy. These are discussed below. 

How can we identify the involved channels without needing a full theoretical model of the harmonic emission? Such identification requires a differential measurement, a measurement which reveals the response of the molecular system to a variation of the laser parameters. This can be achieved by measuring the variation of the harmonic signal with laser ellipticity, $\epsilon$. The measured harmonic signal as a function of laser ellipticity in argon and in SF$_6$ is shown in Fig. \ref{FigEllipticity}(a-b). The harmonic intensity decays exponentially with $\epsilon^2$ \cite{ivanov96} and can be fitted by a Gaussian $I_q(\epsilon)=I_q^0 e^{-\beta_q\epsilon^2}$, where $q$ is the harmonic order and $\beta_q$ the variation with ellipticity. The ellipticity-dependent responses of argon and SF$_6$ are remarkably similar for harmonic 21 whereas that of argon is higher for harmonic 15.

High-order harmonic emission in an elliptical laser field can be described in a quasi-static semi-classical picture,  as follows. First, electrons tunnel out of the molecule with a finite transverse momentum distribution. This distribution is dictated by the molecular orbital structure, filtered through the tunneling process \cite{murray11}. If the electron wavepacket at the exit of the tunnel is spatially localized, then the momentum distribution is broad. The transverse momentum distributions associated with argon and the A and C channels in SF$_6$, calculated using the approach described in \cite{murray11}, are shown in Fig. \ref{FigEllipticity}(c). The distributions have a Gaussian profile, with the same width for argon and the A channel, while the C channel shows a broader distribution. This is consistent with the more localized character of the HOMO-3 (C) rather than the diffuse HOMO-1 (A). During propagation in the continuum, the electron wavepacket 
spreads transversly. The spread of the returning electron wavepacket depends on the time spent in the continuum (and thus on the harmonic order, the lowest harmonics being emitted before 
the highest harmonics \cite{mairesse03}), but also on the initial transverse momentum distribution: the broader the initial transverse momentum distribution, the broader the returning wavepacket. With an elliptically polarized driving field, the orthogonal component of the  field shifts the electron in the lateral direction, leading to a suppression of the recollision probability. Such suppression depends on both the trajectory length and the initial channel-specific transverse momentum. Therefore, the harmonic response $\beta$ as a function of  ellipticity serves as a fingerprint of the molecular orbital, i.e. of the SFI channel. 

The variation  of $\beta_{SF_6}/\beta_{Ar}$ with harmonic order is shown in Fig. 5(d). Normalization to the response of argon suppresses the linear dependence of the ellipticity response  with harmonic order, which simply reflects the change in electron trajectory length. Above H17, the ellipticity responses  of SF$_6$ and argon are remarkably similar, while they differ by $\sim 20 \%$ for H15 and below. Theoretical ellipticity responses  are calculated using Gaussian fits of the distributions shown in Fig. \ref{FigEllipticity}(c) and reported, in Fig. 5(d), upon normalization to the argon response. They are in good agreement with experiment, given the simplicity of the model, and confirms the transition between the A and C channels at H17.

\subsection{Measurement of the relative phase between channels}
\begin{figure}
\begin{center}
\includegraphics[width=0.5\textwidth]{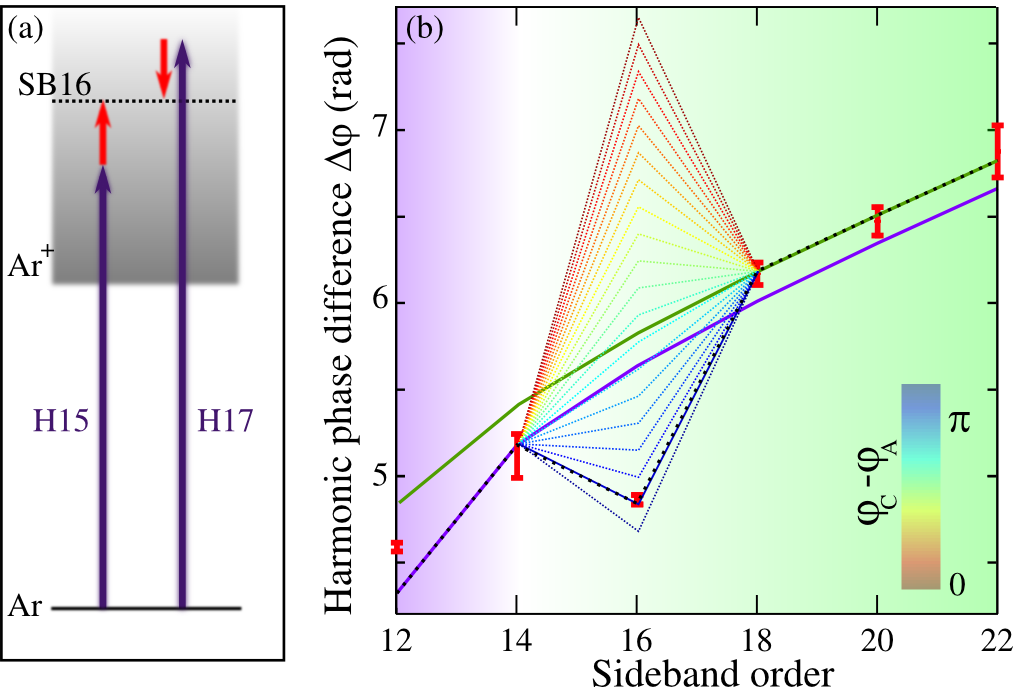}
\caption{Measurement of the relative phase between HHG channels. (a) Principle of the RABBIT method: high harmonics photoionize argon atoms in the presence of an infrared (800 nm) field. Absorption of harmonic 15 and one IR photon, or absorption of harmonic 17 and stimulated emission of one IR photon both lead to  the production of the same sideband (SB16) in the photoelectron spectrum. (b) Phase of the sideband oscillations as a function of the delay between IR and harmonics. The experimental data points (with error bars) and theoretical results (solid lines) from saddle point analysis for the A (green) and C (purple) channels. To account for the response at SB16, we consider a transition from A to C channels between harmonic 15 and 17 (black dots). In this case, the relative phase between the two channels determines the magnitude of the signal  for SB16. The dotted lines are the results of simulations as a function of  relative phases $\varphi_C-\varphi_A$ varying from 0 to $\pi$ by steps of 0.05$\pi$. It 
can be seen that the  experimental data agrees with the simulations for a relative phase  of $0.95\pi$. }
\label{FigRabbit}
\end{center}
\end{figure}

Elliptically-driven HHG in SF$_6$ revealed a transition in the channel dominating the emission process: harmonics 15 and below are produced by the C channel whereas harmonics 17 and above are produced by the A channel. In the following we use attosecond metrology to determine the relative phase between these two channels.

The RABBIT technique \cite{paul01} is an attosecond pulse characterization method which measures the relative phase between consecutive harmonics. RABBIT is based on photoionization of a target gas by high-order harmonics, in the presence of a weak fundamental laser field. The photoelectron spectrum shows main lines, resulting from the absorption of harmonics, and sidebands, corresponding to absorption of one harmonic and absorption or emission of an infrared photon (Fig. \ref{FigRabbit}(a)). Two quantum paths involving two consecutive harmonics lead to the same sideband, which results in an interference process. The sideband intensity thus oscillates as a function of delay between harmonics and the IR. The phase of this oscillation $\Phi^{SB}_q$ encodes the relative phase between the neighbouring harmonics $q+1$ and $q-1$: $\Phi^{SB}_q=\Delta\varphi_{q}+\Delta\phi^{at}_q$, where $\Delta\varphi_
q=\varphi_{q+1}-\varphi_{q-1}$ is the phase difference between harmonics $q+1$ and $q-1$ and $\Delta\phi^{at}_q$ is a scattering phase intrinsic to the target (the atomic phase, i.e. the phase accumulated by the photoionized electron in the atomic scattering potential \cite{paul01}).

In Figure \ref{FigRabbit}(b), we show the measured harmonic phase difference obtained using SF$_6$ as a generating gas and argon as a detecting gas. The harmonic phase difference increases with harmonic order but a clear deviation is observed at SB16, reflecting a specific phase difference between harmonic 15 and 17. A similar behavior was reported in \cite{rothhardt12}. Calculations of the harmonic phase, within the Strong-Field Approximation and using the Saddle Point method, were performed for the two channels expected to dominate the harmonic emission in this energy range: the A (green) and C (purple) channels. The resulting harmonic phase difference shows a similar increase with harmonic order for the two channels, reflecting the temporal dispersion of electron trajectories leading to emission of different harmonics \cite{mairesse03}. The calculations are in good agreement with the experiment for SB 18 to 22, associated with channel A, and for SB 12-14, associated with channel C. We note that a 
transition from the C to A HHG channels between harmonic 15 and 17 would produce a shift in the phase of sideband 16. The electron trajectory leading to a given harmonic order slightly differs between the two channels, and this difference is enough to induce a significant effect on the phase accumulated along the whole trajectory. However, the results from this model do not agree with the experimental data: the calculated $\Delta\varphi$ for SB16 is around 1.3 rad while the measured phase shift is 4.8 rad. This shows that there is an additional dephasing of the harmonic emission between channels A and C which is directly measured by experiment. Therefore, the simulations were repeated with the introduction of an additional variable phase shift $\varphi_C-\varphi_A$ as an adjustable fit parameter (Fig. \ref{FigRabbit} (b)). Excellent agreement with the experiment is obtained when  $\varphi_C-\varphi_A =$ 0.95$\pi$. Importantly, this  phase difference arises from the phases of the recombination matrix elements 
in the HHG process. Its determination illustrates the strength of this interferometric method in revealing the effects of shape resonances, autoionizing states and ionization phases in HHG \cite{haessler13}.

\subsection{Response of different channels to molecular vibrations}
\begin{figure}
\begin{center}
\includegraphics[width=0.5\textwidth]{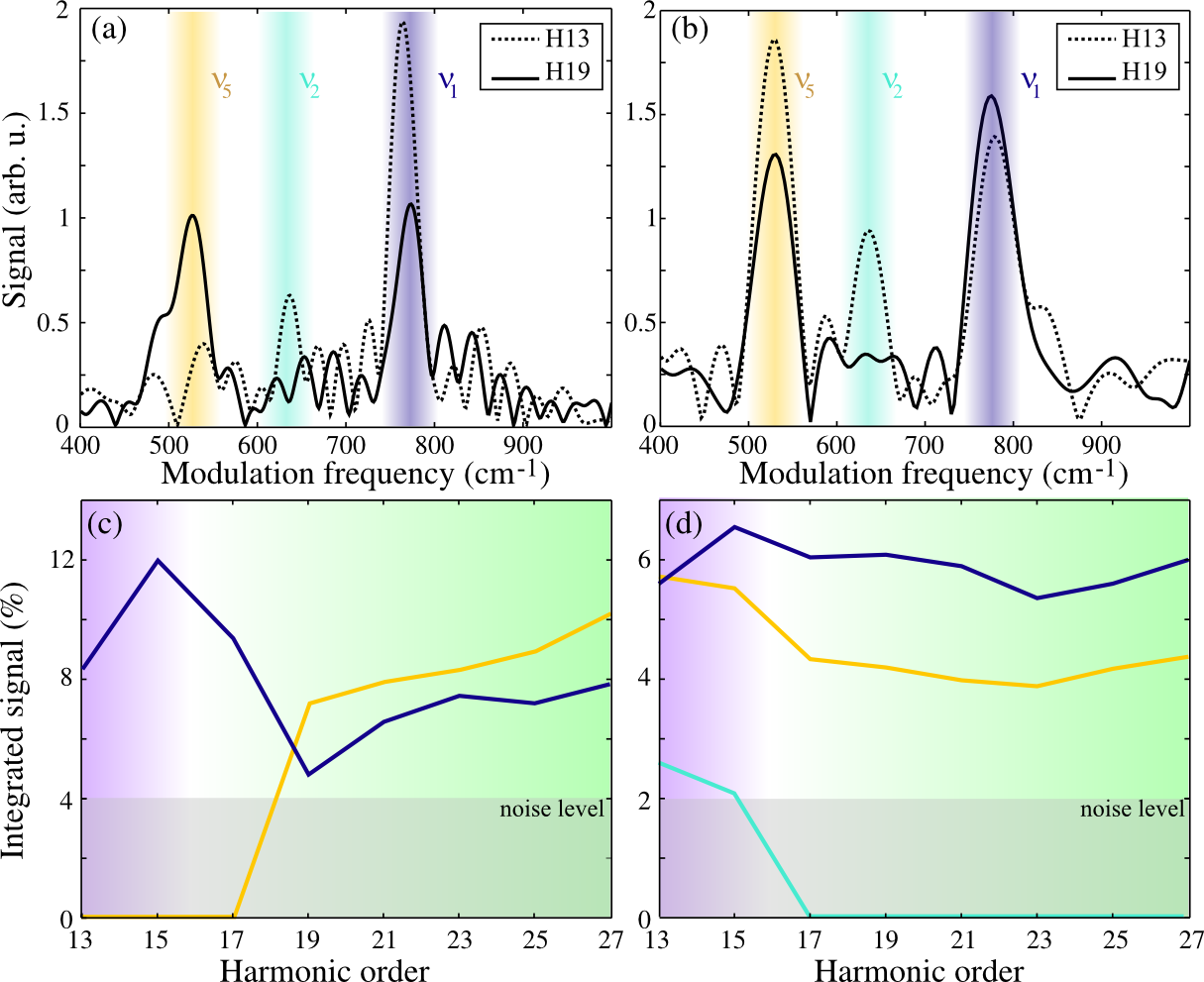}
\caption{The modulation of high harmonic generation by specific vibrations in ground state SF$_6$ molecules. Fourier transform of the intensity (a) and phase (b) modulations with pump-probe delay for harmonic 13 (dots) and 19 (line) showing the contribution of different Raman modes: $\upsilon_5$-T$_{2g}$ (524 cm$^{-1}$, yellow), $\upsilon_2$-E$_{g}$ (643 cm$^{-1}$, green) and $\upsilon_1$-A$_{1g}$ (774 cm$^{-1}$, blue). The contribution of each mode, obtained by integration of the peak in the FFT, is shown as a function of harmonic order in (c) intensity and (d) phase. }
\label{FigDynamique}
\end{center}
\end{figure}

We have identified the different channels operating in HHG from SF$_6$ by manipulating the electron trajectories in the generating mechanism. In the following, we vary  the nuclear degrees of freedom via vibrational excitation, resulting in systematic  modification of all stages of the 'interferometer'. As discussed below, this further   highlights the multi-channel nature of the HHG process.

Small molecular structural changes in the generating medium can induce significant modifications in the high-harmonic emission. For instance, Raman pump - strong-field probe experiments in SF$_6$ showed that the intensity of harmonic 25 to 45  varies by $\sim10\%$ upon molecular vibration \cite{wagner06}, even if the bond length variations remain small ($\sim 1\%$). Molecular vibrations may significantly change both the amplitude as well as the phase of each channel involved in the interferometer. First, changes in molecular geometry are expected to alter the vertical ionization potentials. Such modification influences the relative amplitude as well as the relative phase of each channel. For example, a 65 meV change in I$_p$ results in a 100 mrad phase change for a plateau harmonic. Second, molecular vibrations can lift the electronic degeneracy of molecular orbitals, increasing the number of possible channels. Finally, any distortions of the molecular orbitals associated with these  vibrations may affect 
both the ionization and recombination dipole moments.

We have performed pump-probe experiments using an intense ($\sim 7\times10^{13}$ W.cm$^{-2}$) 800 nm pulse to Raman-pump vibrational modes in the SF$_6$ molecules and produce high harmonics in the excited medium. In addition to conventional experiments in which the harmonic intensity is measured (Fig. \ref{FigDynamique}(a,c)), we have used two-source interferometry to determine the evolution of the harmonic phase with pump-probe delay (Fig. \ref{FigDynamique}(b,d)) \cite{smirnova09,camper14} (see Methods). The harmonic emission is periodically modulated as a function of pump-probe delay. The typical modulation depth of the harmonic signals are  5$\%$ for the intensity and 100 mrad for the harmonic phase. Three of the six normal modes of SF$_6$ are Raman active vibrations: $\upsilon_1$-fully symmetric and strongly active mode $A_{1g}$ with a quantum of 774 cm$^{-1}$ (vibrational period of $\approx$ 43 fs), $\upsilon_2$-the doubly degenerate mode $E_g$ with a quantum of 643 cm$^{-1}$ ($\approx$ 52 fs) and 
finally the triply degenerated $\upsilon_5$-$T_{2g}$ mode with a quantum of 524 cm$^{-1}$ ($\approx$ 63 fs) \cite{boudon04}. A Fourier analysis is performed to determine which vibrational modes modulate the signal (Fig. \ref{FigDynamique}(a-b)).

The weight of each vibrational mode in the intensity and phase modulations is extracted by integrating the corresponding peak and normalizing to the total signal (Fig. \ref{FigDynamique}(c-d)). The results show a clear transition around harmonic 17, consistent with the channel switching measured in static molecules. Below harmonic 17, in the region where the C channel dominates, the harmonic intensity is exclusively modulated by the $\upsilon_1$ mode. Above harmonic 17, the $\upsilon_5$ mode becomes slightly predominant. There is no significant contribution of the $\upsilon_2$ mode except for harmonic 13. Interestingly, the behavior of the harmonic phase is very different but also shows a transition associated to channel switching. The phase emission of C channel does not vary drastically for $\upsilon_1$ and $\upsilon_5$ nuclear dynamics and shows a weak contribution from $\upsilon_2$. The phase of the emission from the A channel seems insensitive to $\upsilon_2$ and dominated by $\upsilon_1$.

The complete calculation of the harmonic emission from vibrating molecules is beyond the scope of this article. As a first step, we evaluated the modifications of the strong field ionization yield resulting from molecular vibrations in the two dominant modes seen in the experiment: $\upsilon_1$ and $\upsilon_5$. For the C channel, ionization is maximized when the molecule is stretched in the $\upsilon_1$ mode, but shows no modulation in the $\upsilon_5$ mode. The ionization from the A channel shows modulations in both modes. This is in remarkable agreement with the measured modulations of the harmonic intensity (Fig. 7(c)), which suggests that when the strong field ionization yield is not modulated, neither is the recombination dipole moment.

\section{Discussion}
Multi-channel high-harmonic generation in polyatomic molecules is a rich and complex process which initiates a broad range of multielectron phenomena. The identification and isolation of the different channels is a major challenge which requires the development and integration of new experimental tools. In this work we introduced and combined several experimental schemes in an advanced spectroscopic study, enabling us to resolve multi-channel SFI in SF$_6$. First, we identified the different SFI channels by performing wavelength-dependent CRATI measurements. Varying  the wavelength addressed the issue of overlapping ATI combs which can occur at specific wavelengths, notably at the 800 nm driving frequency implemented in the majority of this work. We introduced a new scheme to HHG spectroscopy by resolving the spectral dependence of the normalized ellipticity response.  This measure enabled  
isolation of the transverse shape of the electronic wavefunction at the moment of tunnel 
ionization, removing the complexity imposed by the additional steps of the interaction. We made use of  the coherent properties of the process via RABBIT measurements which enabled the extraction of the relative phase between the channels. Such a measurement identified the important contribution of shape and autoionization resonances, hidden in many HHG measurements. Finally, we have made use of the nuclear degrees of freedom - via vibrational dynamics -- as an important control knob which manipulates both the phase and amplitude associated with each channel, advancing our ability to identify the multi-channel nature of the interaction.

We believe that the methods presented here will illuminate the complexity and the rich spectroscopic information obscured in many HHG experiments. For example, coupled with accurate calculations of the multi-channel harmonic emission, our results should facilitate  understanding the role of resonances in the emission process and also the elucidation of the influence of Jahn-Teller distortions in the ion \cite{walters09}.

The clear identification of the SFI channels is the first fundamental step in characterizing time-resolved multielectron processes, the long term goal being to resolve charge migrations or field induced coupling between the channels. We believe that it will be possible to isolate such dynamics by increasing the dimensionality of the experiment -- by scanning the fundamental wavelength of the strong laser field in HHG. Such scanning will scale the interferometer, presented in Fig. 1, enabling the decoupling of the different degrees of freedom involved in the process. Disentangling, for example, the contribution of resonances from dynamic features will be an important tool in HHG spectroscopy. In summary, we believe that the combination of methods presented here constitute an important building block for strong field spectroscopy and will lead to the discoveries of a broad range of multielectron strong field phenomena in complex targets such as polyatomic molecules.

\section{Methods}
\subsection{Channel-resolved Above-Threshold Ionization}
Experiments were performed using a Legend Elite Duo laser system (Coherent Inc.) that delivered 35 fs, 3.1 mJ, 800 nm pulses at 1 kHz (Molecular Photonics laboratory, NRC Ottawa). Laser fundamental pulses were implemented directly in the 800 nm experiments whereas $\sim$40~fs 400 nm pulses were generated through second harmonic generation in thin $\beta$-BaBO$_4$ crystals (150 $\mu$m thick). $\sim$40 fs 480 nm, 532 nm and 590 nm pulses were generated through sum frequency generation of the signal pulses from an optical parametric amplifier (Light Conversion, TOPAS-C) and the fundamental beam in thin $\beta$-BaBO$_4$ crystals (250 $\mu$m thick). Laser  pulses were focused with a f=50 cm spherical mirror onto a continuous, neat expansion of SF$_6$ molecules (expanded through a 50 $\mu$m pinhole $\sim$60 cm from the laser-molecular beam interaction region in a differentially pumped chamber). The molecular beam PEPICO (Photo-Electron Photo-Ion COincidence) spectrometer used to collect the resulting 
photoelectrons and photoions has been described elsewhere \cite{boguslavskiy12}. Briefly, the kinetic energies  of the photoelectrons were resolved by time-of-flight using a wide-bore permanent magnet, magnetic bottle spectrometer. Photoions were collected using a coaxial pulsed-field Wiley-McClaren time-of-flight mass spectrometer. The photoelectron and photoion time-of-flight signals were post-processed using a covariance analysis \cite{frasinski92,Mikosch13} to correlate specific electron signals with the dominant SF$_5^+$ ions. The typical exponential ATI decay in energy has been substracted to enhance the ATI oscillations. This has to be described in the text (or methods).The resulting ATI spectra and experimental intensities were calibrated by argon and xenon ATI spectra recorded as a function of laser intensity.

\subsection{Calculations of strong field ionization rates}
Bound-state neutral and cation wavefunctions were calculated using aug-cc-pVTZ  basis set \cite{augccpvtz}, large-core effective core potentials \cite{stuttgartecp}, and RASSCF wavefunctions with 35 (cation) or 36 (neutral) electrons in 25 orbitals, allowing single and double excitations. For the cation, the active space was optimized for an equally-weighted average of the six lowest states of the cation (~X through ~D). The RASSCF vertical IPs were shifted uniformly by -0.512 eV.  At the equilibrium $O_h$ geometry (R(S-F)=1.575 Angstrom), the resulting cation state  energies are within 0.3 eV of the correlated MCQDPT2(27/28,21)/aug-cc-pVTZ values. 

The magnitude of the vibrational distortions induced by an impulsive Raman pump was determined \cite{impulsiveraman} from normal vibrational modes and dipole polarizabilities,  calculated at the CCSD/aug-cc-pVTZ level \cite{ccsd-polar}. Calculated displacements assumed pump energy flux of $\approx 8.4$ J cm$^{-2}$. Pump and probe polarizations were assumed to be collinear, and were integrated on an order-9 Lebedev grid \cite{lebedevorder9}. RASSCF and MCQDPT2 \cite{mcqdpt2} calculations were performed with GAMESS-US \cite{gamessus}. CCSD calculations used CFOUR \cite{cfour}. Strong-field ionization calculations \cite{spanner-pra,spanner13} used Cartesian product grid with 0.18 Bohr spacing, extending to +/-18 Bohr from the origin, with an absorbing boundary \cite{manolopulous} of 9.4 Bohr at the box edges. The simulation used 14 spatial channels, corresponding to the ~X, ~A, ~B, ~C, and ~D
final states of the cation. Both laser and correlation coupling terms were included. Leap-frog time evolution used a time step of 0.003 atomic units.

\subsection{High-order harmonic spectroscopy}
The high-harmonic spectroscopy experiments were performed using the Aurore laser system at CELIA, which delivers 28 fs, 7 mJ, 800 nm pulses at 1kHz. High harmonics were generated by focusing the pulses with a f=50 cm lens in a continuous jet produced by a 250 $\mu$m nozzle backed by a few hundred mbars of gas. The emitted radiation was analyzed by an extreme ultraviolet spectrometer (1200 mm$^{-1}$ grating, dual Micro Channel Plates and a CCD camera).

To manipulate the laser ellipticity, two broadband zero-order waveplates were inserted in the laser beam: a motorized half waveplate in front of a fixed quarter waveplate. The harmonic signal was recorded as a function of the angle $\alpha$ between the half and quarter waveplate axis  which determines the laser ellipticity $\epsilon=tan(2\alpha)$.

For RABBIT measurements, we used the experimental setup described in \cite{loch11}. Briefly, the harmonic and infrared beams were delayed using a beam-splitting dichroic delay-line. They were focused by a toroidal mirror in a magnetic bottle time-of-flight spectrometer which recorded the electron spectrum produced by photoionizing argon atoms, as a function of delay between XUV and infrared beams.

Phase-resolved pump-probe experiments were performed by using a 0-$\pi$ phase mask \cite{camper14} inserted on the harmonic generating beam, before focusing. This mask created an interference pattern at focus, which resulted in the creation of two harmonic sources spatially separated by $\sim 100$  $\mu$m. In the far field these sources produced an interference patern, similar to Young's slits. A 800 nm pump beam was superimposed on one of the two sources, exciting the molecules by impulsive stimulated Raman scattering. The evolution of the phase of the harmonic emission as a function of pump probe delay was measured by determining the shift of the interference fringes, using a spatial Fourier transform. 

\section{acknowledgements}
We thank R. Lausten (NRC), R. Bouillaud, C. Medina and L. Merzeau for technical assistance. S.P. acknowledges helpful discussions with and suggestions by M. Spanner (NRC). A. Stolow acknowledges NSERC (Canada) for financial support. We acknowledge financial support of the Conseil Regional d'Aquitaine (20091304003 ATTOMOL and COLA 2 No. 2.1.3-09010502), the European Union (Laserlab-Europe II N°228334), the Ambassade de France au Canada, and NSERC (Canada).


\end{document}